\documentclass{iau} 
\usepackage{graphicx}

\title[Magneto-asteroseismology of massive magnetic pulsators] 
{Magneto-asteroseismology of massive magnetic pulsators}

\author[B.\,Buysschaert, C.\,Neiner \& C.\,Aerts]   
{B.\,Buysschaert$^{1, 2}$, C. Neiner$^1$
 \and C.\,Aerts$^{2,3}$}

\affiliation{$^1$LESIA, Observatoire de Paris, PSL Research University, CNRS, \\Sorbonne Universit\'es, UPMC Univ. Paris 06, Univ. Paris Diderot, Sorbonne Paris Cit\'e, \\ 5 place Jules Janssen, F-92195 Meudon, France  \\ email: {\tt bram.buysschaert@obspm.fr} \\[\affilskip]
$^2$Instituut voor Sterrenkunde, KU Leuven, \\Celestijnenlaan 200D, 3001 Leuven, Belgium \\[\affilskip]
$^3$Dept. of Astrophysics, IMAPP, Radboud University Nijmegen, \\6500 GL Nijmegen, the Netherlands}

\pubyear{2016}
\volume{329}  
\jname{The Lives and Death-throes of Massive Stars}
\editors{A.C. Editor, B.D. Editor \& C.E. Editor, eds.}
\begin{document}

\maketitle

\begin{abstract}
Simultaneously and coherently studying the large-scale magnetic field and the stellar pulsations of a massive star provides strong complementary diagnostics suitable for detailed stellar modelling.  This hybrid method is called magneto-asteroseismology and permits the determination of the internal structure and conditions within magnetic massive pulsators, for example the effect of magnetism on non-standard mixing processes.  Here, we overview this technique, its requirements, and list the currently known suitable stars to apply the method.
\end{abstract}

\firstsection 
                        
\section{Introduction}
Thanks to the efforts of large surveys to detect magnetic fields in stars, such as MiMeS \cite[(Wade \etal\,2016)]{2016MNRAS.456....2W}, BinaMIcS \cite[(Alecian \etal\,2015)]{2015IAUS..307..330A}, the BOB campaign \cite[(Morel \etal\,2015)]{2015IAUS..307..342M}, and the BRITE spectropolarimetric survey \cite[(Neiner \etal\,2016)]{2016arXiv161103285N}, the total number of known magnetic massive  stars continues to increase.  More than 55 massive stars are now known to show a clear magnetic signature in their Stokes\,V measurements.  However, the full extent of the effects of such a large-scale magnetic field being present at the stellar surface remains unexplored.  This magnetic field can, for example, lead to abundance inhomogeneities at the stellar surface or confine circumstellar material in a magnetosphere.  Additionally, it is expected that the stellar structure, and thus also the stellar evolution, will be influenced by this surface magnetic field.

Fortunately, some of these magnetic massive stars also host stellar pulsations.  These permit us to probe the internal conditions at multiple layers, enabling a direct comparison with stellar structure and evolution models.  This provides a unique way of studying the impact of magnetism on, for example, the physics of non-standard mixing processes.  Furthermore, magneto-asteroseismology provides strong observational constraints, since information about the stellar surface and environment is retrieved from the magnetic field, whereas information about the interior comes from pulsations.

\section{Magnetic massive stars}
Magnetic fields are only detected for seven to ten percent of all studied massive hot (spectral type OB) stars, and the field occurrence does not depend on the spectral type \cite[(Grunhut \& Neiner\,2015)]{2015IAUS..305...53G}.  Because these magnetic fields seem to be stable over long timescales and their strength does not seem to correlate with known stellar properties, it is assumed that they are of fossil origin and are frozen into the radiative envelope of the stars.  The fields are those of the birth molecular cloud, partly trapped inside the pre-main sequence star during the cloud collapse phase, possibly further enhanced by a dynamo in the early fully convective stellar phase \cite[(Neiner \etal\,2015a)]{2015IAUS..305...61N}.  Typically, the polar field strength ranges from about a hundred Gauss up to several kiloGauss.  However, some weaker fields, below 100\,G, have recently been detected \cite[(e.g. Fossati \etal\,2015)]{2015A+A...574A..20F}.

The stellar magnetic field influences many different regions of the star with various effects. In the deep interior of the star, the field influences the internal mixing of the star and this affects the size of the convective core overshooting region, changing the lifetime of the star by decreasing the amount of fuel for nuclear burning. Magnetic stars can also confine their stellar winds, due to their strong magnetic fields, into a magnetosphere, which slows down the rotational velocity of the star. This process is called magnetic braking and its effect is observed by the low (projected) rotational velocities for strongly magnetic stars. This magnetic braking is essentially an efficient transport of angular momentum \cite[(Mathis \& Zahn\,2005; Maeder \& Meynet\,2014)]{2005A+A...440..653M, 2014ApJ...793..123M}. At the stellar surface, the magnetic fields can create and sustain areas of chemical over- or under-abundances and/or large temperature differences, which are called spots.

\section{Pulsating massive stars}
Depending on the type of pulsations and accounting for differences in global stellar properties, massive pulsators are classified as different types \cite[(see Aerts \etal\,2010 for a monograph on asteroseismology)]{2010aste.book.....A}.  The most massive main-sequence stars (having spectral type O9\,--\,B2) are labeled $\beta$\,Cep pulsators.  They mainly oscillate with low-order pressure modes, having periods of several hours.  Less massive pulsating stars (spectral type B2\,--\,B9) are called Slowly Pulsating B-type (SPB) stars, which pulsate with high-order gravity modes with a period of the order half to a few days.  Both type of pulsations are driven by the heat mechanism, related to the iron opacity bump \cite[(e.g. Dziembowski \etal\,1993)]{1993MNRAS.265..588D}.  Since their theoretical instability regions overlap, hybrid $\beta$\,Cep/SPB pulsators are expected around spectral type B2 and have been observed.  Moreover, the occurrence of stellar pulsations seems to be uncorrelated to the presence of large-scale magnetic fields.  Thanks to the advent of dedicated space-missions, such as MOST \cite[(Walker \etal\,2003)]{2003PASP..115.1023W}, CoRoT \cite[(Baglin \etal\,2006)]{2006cosp...36.3749B}, Kepler/K2 \cite[(Borucki \etal\,2010; Howell \etal\,2014)]{2010Sci...327..977B, 2014PASP..126..398H}, and the BRITE constellation \cite[(Weiss \etal\,2014)]{2014PASP..126..573W}, the number of known pulsating massive stars has drastically increased over the last decade.

To be able to successfully relate the stellar pulsations to the internal properties, by means of comparing them to detailed stellar and seismic models, the geometry of the pulsation mode needs to be known.  This is nearly impossible from white-light photometry, except when rotational splitting is detected and / or for SPB pulsators where regular period patterns permit a direct approach to perform the mode identification.  Therefore, one generally studies the line profile variations (LPVs) seen in metallic absorption lines or the amplitude ratios from multi-color photometry to unravel the mode geometry \cite[(e.g. De Cat \etal\,2005)]{2005A+A...432.1013D}.

\begin{table}[!t]
\tabcolsep=1pt
  \begin{center}
  \caption{Known magnetic massive pulsators, having $N$ detected pulsation modes.}
  \label{tab:Targets}
    {\scriptsize
    \begin{tabular}{|l|l|l|l|c|c|l|l|l|l|}
    \hline 
    & Star 	& $N$ 		& Type	&$P_{\rm rot}$	& $B_{\rm pol}$	& Magnetic			& Binary?		& SpT	& References\\
    &	  	& 	 		& 		& [d]		    & [G]			& characterization	&				&		&			\\
    \hline
1	&	HD\,43317			&	$> 100$	&	both			&$0.90$	&	$\sim 900$	& dip.; $i\in [20, 50]^{\circ}$; $\beta\in [70, 86]^{\circ}$	&		&	B3IV		& (1), (2)	\\    
2	&	$\beta$\,Cen\,Ab		&	$< 17$	& $\beta$\,Cep	&		&	$\sim 250$	&															&	Y	&	+ B1III	& (3), (4)	\\
3	&	$\beta$\,Cep			&	5		& $\beta$\,Cep	&$12.0$	&	$\sim 300$	& dip.; $i\sim60^{\circ}$; $\beta\sim95^{\circ}$				&	Y	&	B0III +	& (5), (6)	\\
4	&	V2052\,Oph			&	3		& $\beta$\,Cep	&$3.64$	&	$\sim 400$	& dip.; $i\sim70^{\circ}$; $\beta\sim35^{\circ}$				&		&	B2IV/V	& (7), (8), (9)\\
5	&	$\beta$\,CMa			&	3		& $\beta$\,Cep	&		&	$ < 30$		&															&		&	B1II/III	& (10), (11), (12)\\
6	&	16\,Peg				&	3		& $\beta$\,Cep	&$1.44$	&	$\sim 500$	& dip.; $i\sim70^{\circ}$; $\beta\sim70^{\circ}$				&		&	B3V		& (13), (14), (15)\\
7	&	$\epsilon$\,Lup\,A	& 'LPV bump'	& $\beta$\,Cep 	&		&	$\sim 600$	&															&	Y	&	B2V +	& (16), (17)\\
	&	$\epsilon$\,Lup\,B	& $> 2$		& $\beta$\,Cep	&		&	$\sim 300$	&															&	Y	&	+ B3V	& (16), (18)\\
8	&	$\xi^1$\,CMa			&	1		& $\beta$\,Cep	&$2.18$	&	$\sim 600$	&															&		&	B1V		& (19), (20)	\\
9	&	HD\,96446			&	1		& $\beta$\,Cep	&$23.4$	&	$\sim 7500$	&															&		&	B2IIIp	& (21), (22), (23)	\\
10	&	$\zeta$\,Cas			&	1?		& SPB			&$5.37$	&	$\sim 150$	& dip.; $i\sim30^{\circ}$; $\beta\sim105^{\circ}$			&		&	B2IV/V	& (24), (25)	\\
11	&	$\sigma$\,Lup		&	1?		& SPB			&$3.09$	&	$\sim 300$	& dip.; $i\sim60^{\circ}$; $\beta\sim90^{\circ}$				&		&	B2III	& (26), (14)	\\
12	&	$\phi$\,Cen			&'LPV bump'	& $\beta$\,Cep	&$1.14$	&	$\sim 900$	&															&		&	B2IV		& (27), (28)	\\
    \hline
    \end{tabular}
    }
 \end{center}
 \vspace{1mm}
 \scriptsize{
 {\it Notes:}\\
  (1): \cite[Briquet \etal\,2013]{2013A+A...557L..16B};  (2) \cite[P{\'a}pics \etal\,2012]{2012A+A...542A..55P};  (3): \cite[Alecian \etal\,2011]{2011A+A...536L...6A}; (4): \cite[Pigulski \etal\,2016]{2016A+A...588A..55P};  (5): \cite[Henrichs \etal\,2013]{2013A+A...555A..46H};  (6): \cite[Telting \etal.\,1997]{1997A+A...322..493T};  (7): \cite[Neiner \etal\, 2012a]{2012A+A...537A.148N};  (8): \cite[Handler \etal\,2012]{2012MNRAS.424.2380H};  (9): \cite[Briquet \etal\,2012]{2012MNRAS.427..483B};  (10): \cite[Fossati \etal\,2015]{2015A+A...574A..20F};    (11): \cite[Mazumdar \etal\,2006]{2006A+A...459..589M};  (12): \cite[Shobbrook \etal\,2006]{2006MNRAS.369..171S};  (13): \cite[Henrichs \etal\,2009]{2009IAUS..259..393H};  (14) \cite[Koen \& Eyer\,2002]{2002MNRAS.331...45K};  (15): \cite[De Cat \etal\,2007]{2007A+A...463..243D};   (16): \cite[Shultz \etal\,2015]{2015MNRAS.454L...1S};  (17): \cite[Schrijvers \etal\,2002]{2002ASPC..259..204S};  (18): \cite[Uytterhoeven \etal\,2005]{2005A+A...440..249U};  (19): \cite[Hubrig \etal\,2006]{2006MNRAS.369L..61H}; (20): \cite[Williams\,1954]{1954PASP...66..200W};  (21): \cite[Borra \& Landstreet\,1979]{1979ApJ...228..809B};  (22): \cite[J{\"a}rvinen \etal\,2017]{2017MNRAS.464L..85J};  (24): \cite[Neiner \etal\,2012b]{2012A+A...546A..44N};  (24): \cite[Briquet \etal\,2016]{2016A+A...587A.126B}; (25): \cite[Neiner \etal\,2003]{2003A+A...406.1019N};  (26): \cite[Henrichs \etal\,2012]{2012A+A...545A.119H}; (27): \cite[Alecian \etal\,2014]{2014A+A...567A..28A}; (28): \cite[Telting \etal\,2006]{2006A+A...452..945T}.}    
\end{table}

\section{Magnetic massive pulsators}
\subsection{Magneto-asteroseismology}
We speak of magneto-asteroseismology when we study magnetic pulsators in a coherent manner, combining both the magnetometric and seismic studies.  Each independent study contains information or constrains observables from a specific layer of the star.  From asteroseismology, we gain information on the density, composition, and chemical mixing in multiple internal layers (depending on the number of studied frequencies).  Additionally, when rotationally split pulsation modes are observed, the internal rotation profile can be retrieved \cite[(e.g. Triana \etal\,2015)]{2015ApJ...810...16T}.  From magnetometry, surface properties are determined, related to the chemical composition, including spots, and the magnetic field, such as its geometry, obliquity, and strength.  Magnetic studies also provide constraints about the wind geometry and the circumstellar environment.  Moreover, the stellar rotation period and the inclination angle towards the observer are also retrieved.

However, the seismic information is only available when two (or more) stellar pulsations have been observed and their mode order and degree have been determined, since one performs a differential study between each probed internal layer.  In addition, the complete magnetic characterisation can only be performed when the rotation period can be determined from the study of the field over the complete rotation period.  At present, only 12 magnetic hot pulsators are known, of which only a few have a fully characterized magnetic field and well studied stellar pulsations (see Table\,\ref{tab:Targets}).  Furthermore, magneto-asteroseismic studies have only been successfully performed for $\beta$\,Cep \cite[(Shibahashi \& Aerts\,2000)]{2000ApJ...531L.143S} and V2052\,Oph \cite[(Briquet \etal\,2012)]{2012MNRAS.427..483B} so far, in part because such a study is observationally demanding.

\subsection{V2052\,Oph}
V2052\,Oph is a magnetic $\beta$\,Cep pulsator, for which 3 stellar pulsation modes have been detected and studied from ground-based spectroscopy and multi-color photometry \cite[(Briquet \etal\,2012, Handler \etal\,2012)]{2012MNRAS.427..483B, 2012MNRAS.424.2380H}.  Its magnetic field was deduced to be dipolar with a strength of $\sim 400$\,G, inclined 35\,$^{\circ}$ to the rotation axis \cite[(Neiner \etal\,2012a)]{2012A+A...537A.148N}.  From detailed seismic modelling, \cite[Briquet \etal\,(2012)]{2012MNRAS.424.2380H} showed that the convective core overshooting region is small, with no extra mixing in spite of the relatively large rotational velocity.  This clearly illustrated the effect of the magnetic field: the inhibition of chemical mixing processes by the magnetic field.  This result is in agreement with theoretical criteria \cite[(e.g. Spruit\,1999; Mathis \& Zahn\,2005)]{1999A+A...349..189S, 2005A+A...440..653M}, which predict that a surface magnetic field of $\sim 70$\,G is sufficient to limit the overshooting in this star.  V2052\,Oph is now considered as a prime example of what magneto-asteroseismology can achieve.

\subsection{Magnetic splitting}
Similar to rotation, the (internal) magnetic field can modify the stellar pulsations by lifting some of its degeneracy \cite[(e.g. Hasan \etal\,2005, and references therein)]{2005A+A...444L..29H}.  Instead of just one pulsation frequency, a multiplet of frequencies is then observed, where the number of frequency peaks is governed by the mode geometry, field geometry and rotation velocity, the size of the constant frequency splitting by the strength of the magnetic field and the rotation velocity, and the amplitude of the individual peaks by the geometry of the magnetic field.  This effect is known as magnetic splitting, and it was proposed as a possible explanation for the observed frequency pattern of $\beta$\,Cep \cite[(Shibahashi \& Aerts\,2000)]{2000ApJ...531L.143S}.  Note, however, that these authors assumed a 6\,d rotation period, instead of the now known 12\,d.  In practice, the magnetic splitting is difficult to observe, because of the very small expected frequency difference between the peaks.  However, it may lead to a wrong mode identification when unaccounted for.  The current best candidate to detect magnetic splitting is  HD\,43317, since this star displays two close frequency patterns \cite[(P{\'a}pics \etal\,2012)]{2012A+A...542A..55P}.

\subsection{Distorted Stokes spectra}
Stellar pulsations can distort spectral absorption lines.  The same effect is also observed in line-averaged spectra, such as those generated by the Least-Squares Deconvolution method using in spectropolarimetric studies; both for Stokes\,I (intensity) and Stokes\,V  (polarization) profiles.  Luckily, this effect seems to be minimal for most spectropolarimetric observations of pulsating magnetic stars, as the exposure times are carefully tailored to remain well below the dominant or shortest pulsation period.  However, in some cases, the pulsation geometry or the overall effect of the LPVs can be so large that LPVs have to be accounted for when determining the longitudinal magnetic field measurements.  Currently, this is done by constructing and fitting surface averaged synthetic spectra to the observations, for which the surface has been distorted by both the stellar pulsations and the magnetic field.  This concept was successfully demonstrated for $\beta$\,Cep, employing the PHOEBE\,2.0 pre-alpha code \cite[(Neiner \etal\,2015b)]{2015IAUS..307..443N}.

\section{Conclusions}
To study magnetic massive pulsators accurately, one needs to account for stellar pulsations when performing the magnetic observations and analysis.  Conversely, the magnetic field might complicate the mode identification of the stellar pulsations.  However, carefully combining both studies, strong complementary observational constraints, e.g. the amount of chemical mixing, the rotation period or the inclination angle, lead to a tightly confined seismic and stellar model.  This permits to study the internal properties of such stars and the effect a large-scale magnetic field has on them.  It is only thanks to recent surveys and dedicated space missions that the sample of suitable targets has become sufficient to start to perform magneto-asteroseismology on a more regular basis.


\begin{thebibliography}{}
\bibitem[Aerts \etal (2010)]{2010aste.book.....A}
{Aerts, C., Christensen-Dalsgaard, J., Kurtz, D.W.} 2010,
\textit{Asteroseismology, Astronomy and Astrophysics Library.~ISBN 978-1-4020-5178-4.~Springer Science+Business Media B.V.}

\bibitem[Alecian \etal (2011)]{2011A+A...536L...6A}
{Alecian, E., Kochukhov, O., Neiner, C., \etal} 2011,
\textit{A\&A}, 536, L6

\bibitem[Alecian \etal (2014)]{2014A+A...567A..28A}
{Alecian, E., Kochukhov, O., Petit, P., \etal} 2014,
\textit{A\&A}, 567, A28

\bibitem[Alecian \etal (2015)]{2015IAUS..307..330A}
{Alecian, E., Neiner, C., Wade, G.A., \etal} 2015,
\textit{IAUS}, 307, 330

\bibitem[Baglin \etal (2006)]{2006cosp...36.3749B}
{Baglin, A., Auvergne, M., Boisnard, L., \etal} 2006,
\textit{COSP}, 36, 3749

\bibitem[Borra \& Landstreet (1979)]{1979ApJ...228..809B}
{Borra, E.F. \& Landstreet, J.D.} 1979,
\textit{ApJ}, 228, 809

\bibitem[Borucki \etal (2010)]{2010Sci...327..977B}
{Borucki, W.J., Koch, D., Basri, G., \etal} 2010,
\textit{Science}, 327, 977

\bibitem[Briquet \etal (2012)]{2012MNRAS.427..483B}
{Briquet, M., Neiner, C., Aerts, C., \etal} 2012,
\textit{MNRAS}, 427, 483

\bibitem[Briquet \etal (2013)]{2013A+A...557L..16B}
{Briquet, M., Neiner, C., Leroy, B., \etal} 2013,
\textit{A\&A}, 557, L16

\bibitem[Briquet \etal (2016)]{2016A+A...587A.126B}
{Briquet, M., Neiner, C., Petit, P., \etal} 2016,
\textit{A\&A}, 587, A126

\bibitem[De Cat \etal (2005)]{2005A+A...432.1013D}
{De Cat, P., Briquet, M., Daszy{\'n}ska-Daszkiewicz, J., \etal} 2005,
\textit{A\&A}, 4, 1013

\bibitem[De Cat \etal (2007)]{2007A+A...463..243D}
{De Cat, P., Briquet, M., Aerts, C., \etal} 2007,
\textit{A\&A}, 463, 243

\bibitem[Dziembowski \etal (2012)]{1993MNRAS.265..588D}
{Dziembowski, W.A., Moskalik, P., Pamyatnykh, A.A.} 1993,
\textit{MNRAS}, 265, 588

\bibitem[Fossati \etal (2015)]{2015A+A...574A..20F}
{Fossati, L., Castro, N., Morel, T., \etal} 2015,
\textit{A\&A}, 574, A20

\bibitem[Grunhut \& Neiner (2015)]{2015IAUS..305...53G}
{Grunhut, J.H. \& Neiner, C.} 2015,
\textit{IAUS}, 305, 53

\bibitem[Handler \etal (2012)]{2012MNRAS.424.2380H}
{Handler, G., Shobbrook, R.R., Uytterhoeven, K., \etal} 2012,
\textit{MNRAS}, 424, 2380

\bibitem[Henrichs \etal (2009)]{2009IAUS..259..393H}
{Henrichs, H.F., Neiner, C., Schnerr, R.S., \etal} 2009,
\textit{IAUS}, 259, 393

\bibitem[Henrichs \etal (2012)]{2012A+A...545A.119H}
{Henrichs, H.F., Kolenberg, K., Plaggenborg, B., \etal} 2012,
\textit{A\&A}, 545, A119

\bibitem[Henrichs \etal (2013)]{2013A+A...555A..46H}
{Henrichs, H.F., de Jong, J.A., Verdugo, E., \etal} 2013,
\textit{A\&A}, 555, A46

\bibitem[Hasan \etal (2005)]{2005A+A...444L..29H}
{Hasan, S.S., Zahn, J.P., Christensen-Dalsgaard, J.} 2005,
\textit{A\&A}, 444, L29

\bibitem[Howell \etal (2014)]{2014PASP..126..398H}
{Howell, S.B., Sobeck, C., Haas, M., \etal} 2014,
\textit{PASP}, 126, 398

\bibitem[Hubrig \etal (2006))]{2006MNRAS.369L..61H}
{Hubrig, S., Briquet, M., Sch{\"o}ller, M., \etal} 2006,
\textit{MNRAS}, 369, L61

\bibitem[J{\"a}rvinen \etal (2017)]{2017MNRAS.464L..85J}
{J{\"a}rvinen, S.P., Hubrig, S., Ilyin, I.} 2017,
\textit{MNRAS}, 464, L85

\bibitem[Koen \& Eyer (2002))]{2002MNRAS.331...45K}
{Koen,  C. \& Eyer, L.} 2002,
\textit{MNRAS}, 331, 45

\bibitem[Maeder \& Meynet (2014)]{2014ApJ...793..123M}
{Maeder, A. \& Meynet, G.} 2014,
\textit{ApJ}, 793, 123

\bibitem[Mathis \& Zahn (2005)]{2005A+A...440..653M}
{Mathis, S. \& Zahn, J.P.} 2005,
\textit{A\&A}, 440, 653

\bibitem[Mazumdar \etal (2006)]{2006A+A...459..589M}
{Mazumdar, A., Briquet, M., Desmet, M., Aerts. C.} 2006,
\textit{A\&A}, 459, 589

\bibitem[Morel \etal (2015)]{2015IAUS..307..342M}
{Morel, T., Castro, N., Fossati, L., \etal} 2015,
\textit{IAUS}, 307, 342

\bibitem[Neiner \etal (2003)]{2003A+A...406.1019N}
{Neiner, C., Geers, V.C., Henrichs, H.F., \etal} 2003,
\textit{A\&A}, 406, 1019

\bibitem[Neiner \etal (2012a)]{2012A+A...537A.148N}
{Neiner, C., Alecian, M., Briquet, M., \etal} 2012a,
\textit{A\&A}, 537, A148

\bibitem[Neiner \etal (2012b)]{2012A+A...546A..44N}
{Neiner, C., Landstreet, J.D., Alecian, E., \etal} 2012b,
\textit{A\&A}, 546, 44

\bibitem[Neiner \etal (2015a)]{2015IAUS..305...61N}
{Neiner, C., Mathis, S., Alecian, E., \etal} 2015a,
\textit{IAUS}, 305, 61

\bibitem[Neiner \etal (2015b)]{2015IAUS..307..443N}
{Neiner, C., Briquet, M., Mathis, S., Degroote, P.} 2015b,
\textit{IAUS}, 307, 443

\bibitem[Neiner \etal (2016)]{2016arXiv161103285N}
{Neiner, C., Wade, G., Marsden, S., Blaz{\`e}re, A.} 2016,
\textit{ArXiv e-prints}, 1611.03285

\bibitem[P{\'a}pics \etal (2012)]{2012A+A...542A..55P}
{P{\'a}pics, P.I., Briquet, M., Baglin, A., \etal} 2012,
\textit{A\&A}, 542, A55

\bibitem[Pigulski \etal (2016)]{2016A+A...588A..55P}
{Pigulski, A., Cugier, H., Popowicz, A., \etal} 2016,
\textit{A\&A}, 588, 55

\bibitem[Schrijvers \etal (2002)]{2002ASPC..259..204S}
{Schrijvers, C., Telting, J.H., De Ridder, J.} 2002,
\textit{ASPC}, 259, 204

\bibitem[Shibahashi \& Aerts (2000)]{2000ApJ...531L.143S}
{Shibahashi, H. \& Aerts, C.} 2000,
\textit{ApJ}, 531, L143

\bibitem[Shobbrook \etal (2006)]{2006MNRAS.369..171S}
{Shobbrook, R.R., Handler, G., Lorenz, D., Mogorosi. D} 2006,
\textit{MNRAS}, 369, 171

\bibitem[Shultz \etal (2015))]{2015MNRAS.454L...1S}
{Shultz, M., Wade, G.A., Alecian, E., BinaMIcS collaboration} 2015,
\textit{MNRAS}, 454, L1

\bibitem[Spruit (1999)]{1999A+A...349..189S}
{Spruit, H.C.} 1999,
\textit{A\&A}, 349, 189

\bibitem[Telting \etal (1997)]{1997A+A...322..493T}
{Telting, J.H., Aerts, C., Mathias, P.} 1997,
\textit{A\&A}, 322, 493

\bibitem[Telting \etal (2006)]{2006A+A...452..945T}
{Telting, J.H., Schrijvers, C., Ilyin, I.V., \etal} 2006,
\textit{A\&A}, 452, 945

\bibitem[Triana \etal (2015)]{2015ApJ...810...16T}
{Triana, S.A., Moravveji, E., P{\'a}pics, P.I, \etal} 2015,
\textit{ApJ}, 810, 16

\bibitem[Uytterhoeven \etal (2005)]{2005A+A...440..249U}
{Uytterhoeven, K., Harmanec, P., Telting, J.H., Aerts, C.} 2005,
\textit{A\&A}, 440, 249

\bibitem[Wade \etal (2016)]{2016MNRAS.456....2W}
{Wade, G.A., Neiner, C., Alecian, E., \etal} 2016,
\textit{MNRAS}, 456, 2

\bibitem[Walker \etal (2003)]{2003PASP..115.1023W}
{Walker, G., Matthews, J., Kuschnig, R., \etal} 2003,
\textit{PASP}, 115, 1023

\bibitem[Weiss \etal (2014)]{2014PASP..126..573W}
{Weiss, W.W., Rucinski, S.M., Moffat, A.F.J., \etal} 2014,
\textit{PASP}, 126, 573

\bibitem[Williams (1954)]{1954PASP...66..200W}
{Williams, A.D.} 1954,
\textit{PASP}, 66, 200

\end{thebibliography}
\end{document}